# iit perspektive

Stefan G. Weber

## Alltagstaugliche Biometrie
### Entwicklungen, Herausforderungen und Chancen

Neben zahlreichen Errungenschaften hat die Digitalisierung unserer Alltagswelt auch die Notwendigkeit mit sich gebracht, alltagstaugliche IT-Sicherheitsmechanismen bereitzustellen. Schon seit längerer Zeit wird kontrovers diskutiert, ob eine automatisierte Erkennung menschlicher Charakteristika – kurz: Biometrie – dazu beiträgt, die (nicht nur digitale) Souveränität des einzelnen Nutzers zu verbessern, oder ob sie diese fundamental untergräbt. Dieser Artikel fasst Hintergründe zusammen und zeigt auf, aus welcher Fülle biometrischer Verfahren zu diesem Zwecke geschöpft werden kann – und welchen Herausforderungen dabei zu begegnen ist. Den Chancen des Einsatzes biometrischer Verfahren, insbesondere in Verbindung mit mobilen Endgeräten, werden auch sich abzeichnende Risiken gegenübergestellt.

### Einleitung

Jeder Einzelne von uns interagiert täglich mit einer Vielzahl von vernetzten digitalen Systemen, sowohl bewusst als auch unbewusst. Die enge Verzahnung realer und digitaler Abläufe hat nunmehr zu der paradoxen Situation geführt, dass die prinzipielle Schutzbedürftigkeit der zirkulierenden digitalen Informationen vielfach festgestellt wurde, der Einsatz von Schutzmechanismen jedoch nach wie vor ausgeklammert wird. Die Gründe hierfür sind u. a. in der oft mangelnden Alltagstauglichkeit verfügbarer IT-Sicherheitslösungen zu finden.[1] So stellt die zuverlässige Eingabe eines komplexen Passwortes insbesondere in mobilen Anwendungsszenarien ein großes Problem dar.

In diesem Spannungsfeld stellt sich somit immer häufiger die Frage, welche Ansätze sich prinzipiell eignen könnten, um praktikable und intuitiv zu nutzende Schutzmechanismen zu realisieren.

Biometrische Verfahren rücken zusehends in den Fokus der Aufmerksamkeit, und das nicht zuletzt durch ihre Integration in aktuelle Smartphone-Generationen. Damit haben einzelne Entwicklungslinien bereits den Weg in den Massenmarkt geschafft. Je nach Sichtweise wird dies nur als ein weiterer kleiner Schritt in einer langen Kette betrachtet oder aber als ein großer Schritt für eine ganze Industrie gewertet.[2] Der Einsatz von Biometrie ist in unserer Gesellschaft jedoch noch häufig mit Akzeptanzproblemen vorbelastet. Zumeist wird der Begriff Biometrie pauschal mit Überwachungstechnologien gleichgesetzt und somit als generelle Bedrohungen für die Privatsphäre dargestellt[3], die insbesondere mit einer Strafverfolgung assoziiert wird. Vor allem die Nutzung von Überwachungskamerabildern zur Erkennung von Straftätern anhand ihrer Gesichter ist dabei ein häufig präsentes Szenario in den Medien.

Allseits bekannt ist, dass wir heute schon in unserem Alltag oft und sorglos zahlreiche digitale Spuren hinterlassen, nicht zuletzt durch unsere permanent in die Mobilfunknetze eingebuchten persönlichen Endgeräte. Welche der uns umgebenden Technologien dabei eine tatsächliche Bedrohung für die Privatsphäre darstellen und auch einer Massenüberwachung Tür und Tor öffnen, bleibt angesichts gewachsener Komplexitäten dabei jedoch allzu oft unklar. Auch ist die Nutzung eines weiteren biometrischen Verfahrens – und zwar das der Fingerabdruckerkennung – durch scheinbar mit wenig Aufwand zu manipulierende Sensoren, durch die Nähe zu forensischen Untersuchungsmethoden oder das noch deutlich drastischer ausfallende Schreckensszenario des abgeschnittenen Fingers stigmatisiert.

Es ist jedoch an der Zeit, altbekannte Vorurteile zu überwinden und die Chancen der Technologie in den Fokus der Aufmerksamkeit zu stellen. Der Einsatz von Biometrie hat das Potenzial, große Vorteile im Bereich der Benutzbarkeit von Schutzmechanismen sowie eine starke Personenbindung innerhalb digitaler und realer Abläufe mit sich zu bringen. Somit lässt sich beispielsweise auch die Gefahr des Missbrauchs von Zugriffsberechtigungen stark senken, wodurch im Nachgang auch vorteilhafte Risikomanagementstrategien implementierbar werden. Um dies zu ermöglichen, ist jedoch noch einer Reihe von Herausforderungen zu begegnen.

## Biometrie: Begrifflichkeit und Abgrenzung

Der Begriff Biometrie ist abgeleitet von den griechischen Begriffen „bios" für „Leben" sowie „metron" für „messen". Unter Biometrie werden demnach Verfahren verstanden, die auf einer Messung einzigartiger menschlicher Charakteristika basieren und so eine eindeutige Erkennung eines Individuums ermöglichen.

Im Alltag sind bereits vielfältige Abläufe bekannt, die auf einer Verwendung charakteristischer Merkmale von Individuen beruhen:

▸ Im Rahmen einer Passkontrolle am Flughafen wird das Erscheinungsbild mit den Ausweisdaten abgeglichen, z. B. hinsichtlich Körpergröße, Gesicht und Augenfarbe.
▸ Unterschriften hängen von der Handschrift eines Individuums ab, sie werden z. B. beim Bezahlen mit einer Referenzunterschrift auf der Rückseite einer Kreditkarte verglichen.

Biometrie im engeren Sinne bezieht sich auf eine automatisierte Erfassung und Auswertung individueller Merkmale, bei der die Erkennungsleistung durch Computertechnologien übernommen wird. Grundsätzlich könnten zur biometrischen Erkennung jedoch auch biologische bzw. anatomische Eigenschaften (wie beispielsweise die Strukturen eines Gesichts, einer Iris oder eines Fingers) und verhaltensabhängige Eigenschaften (z. B. das Tippverhalten, die Stimme oder die Gangart) herangezogen bzw. ausgewertet werden.

Grundsätzlich lassen sich zwei Klassen biometrischer Verfahren unterscheiden:

▸ die **Identifizierung** auf Basis biometrischer Merkmale sowie
▸ die **Verifizierung** einer Identitätsbehauptung auf Basis biometrischer Merkmale.

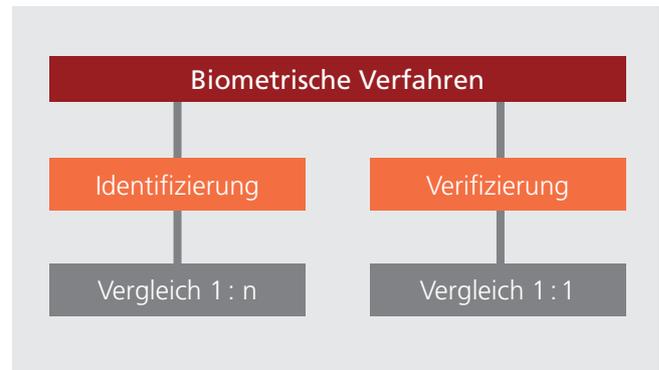

Abb. 1: Abgrenzung Identifizierung / Verifizierung

Bei der Identifizierung findet ein Abgleich eines aktuell erhobenen Messwertes mit vielen vorab gespeicherten Referenzwerten statt. Bekannte Beispiele für die biometrische Identifizierung sind die Nutzung von Fingerabdrücken im Rahmen strafrechtlicher Ermittlungen oder auch die Videoüberwachung öffentlicher Räume. So könnten beispielsweise die Gesichter von Passanten mit denen bekannter Straftäter abgeglichen werden, die in zentralen Datenbanken hinterlegt sind.

Eine Bedrohung der Privatsphäre ergibt sich in biometrischen Systemen zuallererst aus der Speicherung und Auswertung biometrischer bzw. personenbezogener Daten auf Computersystemen außerhalb des Kontrollbereichs des Nutzers. Dadurch kann eine Verkettung sensibler Informationen über Anwendungsgrenzen hinweg vorgenommen werden, anhand derer sich in einem zweiten Schritt weitreichende Benutzerprofile erstellen lassen. Durch die Messung physiologischer Charakteristika besteht zudem die Gefahr, dass unerlaubt weitreichende Schlussfolgerungen über den Gesundheitszustand der „gemessenen Menschen"[4] gezogen werden könnten.

Die Identifizierung stellt jedoch nur einen Verwendungszweck im Rahmen biometrischer Verfahren dar. Nichtsdestotrotz ist die Gleichsetzung „Biometrie = Überwachungstechnik" ein Grund dafür, dass zunächst bei jedwedem Einsatz von Biometrie auch Akzeptanzprobleme zu überwinden sind.

Innerhalb der zweiten Klasse von Verfahren, der sogenannten biometrischen Verifizierung, findet hingegen lediglich ein Abgleich mit einem hinterlegten Referenzdatensatz statt. Dies entspricht der Nutzung eines Passwortes oder einer PIN zur Authentifizierung – hier weist ein Nutzer lediglich die von ihm behauptete Identität nach. Eine biometrische Verifizierung stellt somit eine prinzipielle Alternative zur Authentifizierung auf Basis von Wissen oder Besitz (z. B. mit einer Smartcard) dar.

---

4 Dapp, T. F. (2012): Der vermessene Mensch – biometrische Erkennungsverfahren und mobile Internetdienste, DB Research.

Die folgenden Ausführungen beziehen sich auf die Verwendung biometrischer Verfahren zum Zweck der Verifizierung und damit zur Authentifizierung.

## Wie funktionieren biometrische Vergleiche?

Die Sicherheit eines Passwortes im Sinne der Erratbarkeit hängt von der Länge und der Komplexität der gewählten Zeichenkette ab. Durch komplexe Passwörter werden aber möglicherweise nicht nur Fremde ausgeschlossen, auch die Nutzerfreundlichkeit nimmt ab. Je länger eine Zeichenkette ist, desto schwerer fällt es, sie korrekt zu erinnern und fehlerfrei einzugeben. Sicherheit und Benutzbarkeit stehen somit in einem prinzipiellen Zielkonflikt zueinander. Ein ideales System ist dadurch gekennzeichnet, dass sowohl die Überwindungssicherheit als auch die Benutzbarkeit besonders hoch ausfallen. Traditionelle Authentifizierungsverfahren sind (noch) nicht in der Lage, dies zu leisten. Der Zielkonflikt trifft auch auf biometrische Verfahren zu, jedoch gibt es durch die Kombination verschiedener Verfahren die Möglichkeit, sich dem idealen Verfahren anzunähern.

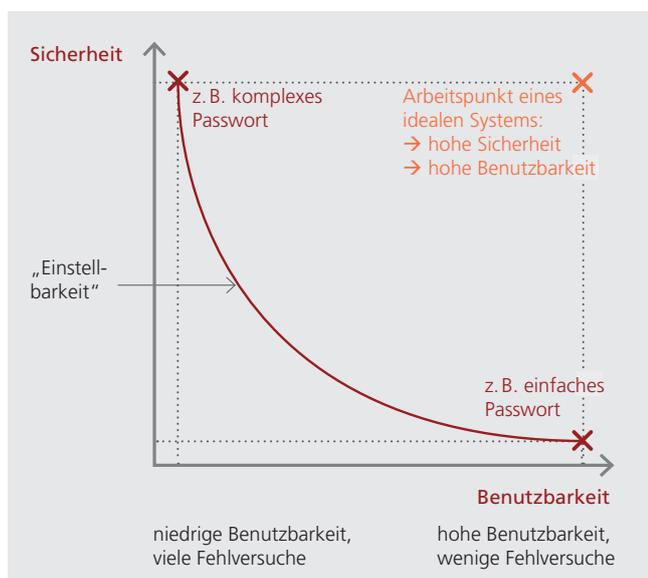

Abb. 2: Tradeoff

Der Vergleichsvorgang, der bei biometrischen Verfahren stattfindet, ist im Unterschied zu wissensbasierten Verfahren kein absolut exakter, sondern ein unscharfer Vergleich. So wie auch verschiedene Unterschriften derselben Person nicht vollständig übereinstimmen, treten bei mehrfachen Messungen eines biometrischen Merkmals regelmäßig Unterschiede auf. Ebenso wie menschliches Sicherheitspersonal am Flughafen hat auch ein automatisiertes biometrisches Verfahren eine gewisse Unschärfe im Abgleich von aktuell erhobenem Messwert und gespeichertem Referenzwert zu berücksichtigen. Während ein Mensch hierzu beispielsweise das aktuelle Erscheinungsbild mit dem Pass abgleicht und dabei auch die Körpergröße abschätzt und die Augenfarbe beurteilt, basiert ein technischer Vergleich hingegen auf einem formalisierten Mustererkennungsvorgang.

Ein gängiger Ansatz besteht darin, aus biometrischen Daten zunächst individuelle Merkmale zu extrahieren. Im Falle der Nutzung von Fingerabdrücken sind dies beispielsweise Abzweigungen, auffällige Schleifen oder Endungen in den Rillen des Fingerabdrucks, die sogenannten Minutien. Bei einem Vergleich zum Zweck der Verifizierung wird anschließend eine hinreichende Übereinstimmung zwischen den Merkmalsmengen geprüft. Die geforderte Übereinstimmung bzw. Überlappung entspricht dabei einer geforderten Passwortlänge und gibt somit die Überwindungssicherheit an. Dies bedeutet, dass die Sicherheit eines biometrisches Verfahrens per se auch von der Konfiguration des Systems abhängig ist. Jedes biometrische System besitzt also eine sogenannte Arbeitslinie, welche die Konfigurierbarkeit bzw. „Einstellbarkeit" vorgibt. Auf dieser Linie ist ein Arbeitspunkt als Parameter auszuwählen. Auch hier findet eine Abwägung zwischen Benutzbarkeit und Sicherheit statt, je restriktiver ein System konfiguriert ist (d. h. je höher das Sicherheitsniveau gewählt wurde), desto mehr Eingabeversuche sind gegebenenfalls notwendig. Prinzipiell können biometrische Systeme im Vergleich zu Passwörtern ein höheres Grundsicherheitsniveau bieten.[5] Gerade hinsichtlich eines Alltagsgebrauchs muss jedoch sichergestellt werden, dass notwendige Sicherheitsstufen nicht unterschritten werden. Dies könnte geschehen, indem beispielsweise von Herstellerseite Parameter vorgegeben werden, welche die Endnutzer vor Fehlversuchen bewahren sollen – und somit die Sicherheit stark herabsetzen. Um die Stabilität und auch Vergleichbarkeit verschiedenster biometrischer Ansätze hinsichtlich der tatsächlich erreichten Sicherheit zu gewährleisten, sind jedoch noch weitere Forschungs- und auch Standardisierungsbemühungen notwendig.

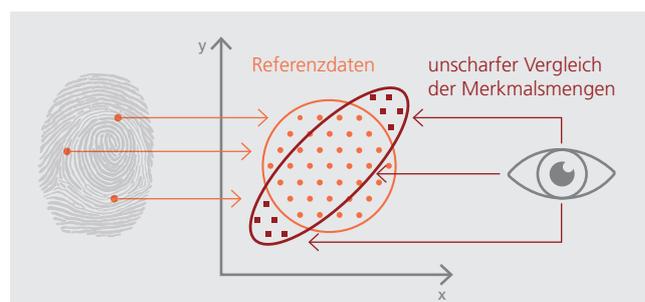

Abb. 3: Unscharfer Vergleich, verschiedene Modalitäten

---

5 Jakobsson, M., Taveau, S. (2012): The Case for Replacing Passwords with Biometrics. In: In Mobile Security Technologies (MoST).

### Wie lassen sich biometrische Daten erfassen?

Damit ein biometrischer Vergleich stattfinden kann, müssen zunächst Eingabedaten erfasst werden. Die entsprechenden technischen Vorrichtungen werden als Sensoren bezeichnet. Biometrische Sensoren können in den verschiedensten Realisierungsformen vorliegen, je nachdem, welche charakteristischen Eigenschaften gemessen werden sollen. Die Erfassung eines Fingerabdrucks erfordert dabei andere technische Vorrichtungen als beispielsweise die Gesichts- oder Gangerkennung. Auch ist es möglich, bestehende Eingabemechanismen als biometrische Sensoren heranzuziehen. So könnte beispielsweise das individuelle Tippverhalten eines Menschen über eine bestehende Computer-Tastatur erfasst werden. Die Nutzung „eingebauter Sensoren" stellt ein aktives Forschungsfeld dar, das u. a. die Herausforderung zu adressieren hat, „stabile" Eingabedaten bereitzustellen.

Eine weitere zentrale Sicherheitseigenschaft biometrischer Verfahren ist die sogenannte Lebenderkennung – allgemeiner ausgedrückt: die Manipulations- bzw. Fälschungserkennung. Sie soll sicherstellen, dass nur legitim eingegebene biometrische Daten als solche akzeptiert werden. Im Falle der Fingerabdruckerkennung sind beispielsweise gelungene Überwindungsversuche mit Fingerattrappen bekannt. Zu diesem Zweck misst ein Fingersensor idealerweise nicht nur die Beschaffenheit der obersten Hautschicht, sondern auch die subepidermalen Schichten. Dieses Gewebe unterscheidet sich erheblich von Ersatzkunststoffen und auch von abgestorbenem Gewebe – ein abgeschnittener Finger würde somit nicht die gewünschte Wirkung erzeugen. Gerade in Bezug auf die Funktionalität der Lebenderkennung in der alltäglichen Praxis muss jedoch festgestellt werden, dass Sicherheitsversprechen der Hersteller zum Teil nicht eingehalten werden. Eine zuverlässige Manipulationserkennung zu implementieren stellt hierbei weiterhin eine große Herausforderung für die Forschung dar.

Im Gegensatz zu Passwörtern besitzen Menschen nur eine sehr begrenzte Anzahl biometrisch nutzbarer Eigenschaften. Als Passwörter können (fast) beliebig komplexe Zeichenketten dienen, die Anzahl verfügbarer Finger ist hingegen begrenzt.[6]

Eine parametrisierte Sensorerfassung trägt zur Erneuerbarkeit biometrischer Referenzdaten bei, sollten Datensätze kompromittiert worden sein. Dies trägt ebenfalls zur Verbesserung der Nutzerakzeptanz bei. Eine weitere große Herausforderung liegt daher in der Entwicklung von Sensoren, die biometrische Daten bereits während der Erfassung geeignet abändern bzw. parametrisieren, ohne die Qualität der Vergleichsvorgänge zu beeinträchtigen.

### Aktuelle Trends: Biometrische Verfahren in mobilen Anwendungsbereichen

Während biometrische Verfahren lange Zeit nur einen kommerziellen Nischenmarkt bedienten, sind sie durch ihre Integration in neue Generationen mobiler Endgeräte auch im Endkundenmarkt angekommen.

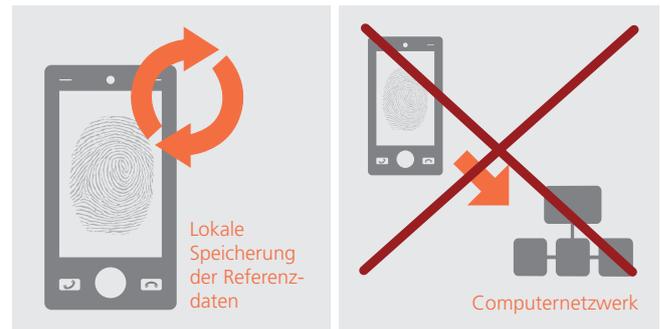

Abb. 4: Lokale Speicherung

Aus Datenschutzsicht stellt sich die dringliche Frage des verantwortungsvollen Umgangs mit biometrischen Referenzdaten. Solange eine Speicherung ausschließlich lokal, also nur auf dem mobilen Endgerät, stattfindet, liegen diese im direkten Kontrollbereich des Nutzers. Ohne eine sorgfältige Analyse des Systemverhaltens kann zunächst jedoch nicht ausgeschlossen werden, dass biometrische Daten auch unbefugt an weitere Stellen bzw. Server weitergeleitet werden. Eine vertrauensbildende Maßnahme könnte deshalb darin bestehen, ein „Datenschutzgütesiegel" – und damit verbunden eine Zertifizierung nach anerkannten Standards – zur Norm zu machen. Ein solch bestandener „digitaler Crashtest" könnte in allgemeinverständlicher Weise den vertrauenswürdigen Umgang mit personenbezogenen biometrischen Daten darlegen.

Eine weiterführende Herausforderung besteht allerdings auch darin, geeignete Hardware- und Softwareplattformen für mobile Endgeräte zu entwickeln. Angelehnt an die derzeit höchsten Industriestandards im Bereich der IT-Sicherheit stellt ein sogenanntes „Secure Element" einen vertrauenswürdigen Lösungsansatz dar. Hierbei handelt es sich um einen manipulationsresistenten Chip, der vom Rest des Betriebssystems des mobilen Endgerätes separiert ist. Findet eine biometrische Verifizierung ausschließlich innerhalb dieses Subsystems statt, so stellt auch eine Kompromittierung des Smartphones keine ernste Bedrohung für die biometrischen Daten dar. Konsequent weitergedacht würde eine derartige Modularisierung aus Datenschutzzwecken zu einem zusätzlichen Endgerät führen, das mit dem Smartphone idealerweise über eine drahtlose

---

6   Kalla, C., Schuch, P. (2013): Sicherheit in der Fingerabdruck-Identifikation. In: Datenschutz und Datensicherheit, Vol. 37 (6), S. 352–357.

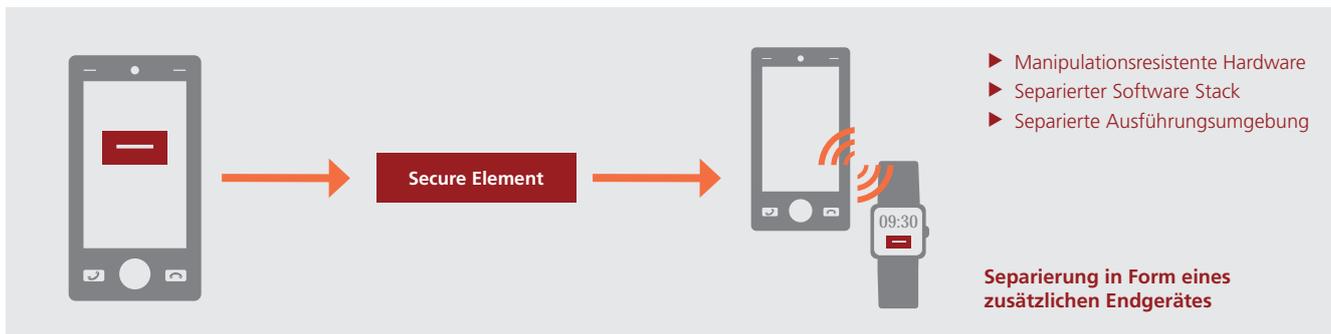

Abb. 5: Sicherheit mobiler Biometrie

Schnittstelle gekoppelt ist. Ein attraktives Design beispielsweise in Form einer Uhr oder eines Schmuckstückes kann zusätzlich dazu beitragen, dass Sicherheitsprodukte zu einem alltäglichen Begleiter werden. Dieses Beispiel soll andeuten, dass eine übergreifende Betrachtungsweise, welche sowohl Datenschutz-, Sicherheits- als auch Designaspekte berücksichtigt, zielführend ist, um die notwendige Akzeptanz für biometrische Authentifizierungslösungen zu erreichen.

## Multimodale und kontinuierliche Biometrie

Neben der Nutzung einzelner biometrischer Charakteristika bietet sich auch eine Kombination verschiedener Modalitäten an. So könnte etwa eine Fingerabdruckerkennung mit einer Iris-Erkennung kombiniert werden. Derzeit werden beispielsweise Verfahren entwickelt, die eine biometrische Verifizierung über die Art und Weise, wie sich eine Person bewegt[7], ermöglichen soll. Eine solche Gangerkennung stellt einen kontinuierlichen Mechanismus bzw. eine kontinuierliche Bindung des Endgerätes an den Nutzer dar. Auf diese Weise ist es möglich, sowohl die Einfachheit der Benutzung als auch die Überwin dungssicherheit zu erhöhen (somit findet eine Annäherung an das skizzierte ideale System statt, vgl. hierzu auch die Abb. 1 zum Tradeoff auf S. 3). Der Besitzer könnte beispielsweise das Endgerät aus seiner Hosentasche ziehen und durch ein kurzes Blinzeln, also eine Iris-Erkennung, ein zur Entsperrung hinreichendes Sicherheitsniveau erreichen. Die Umsetzung dieser Vision erfordert jedoch noch weitere Forschungsbemühungen im Bereich der Fusion biometrischer Merkmale, insbesondere auf mobilen Endgeräten, sowie eine Berücksichtigung weiterer Faktoren der Mensch-Computer-Interaktion.

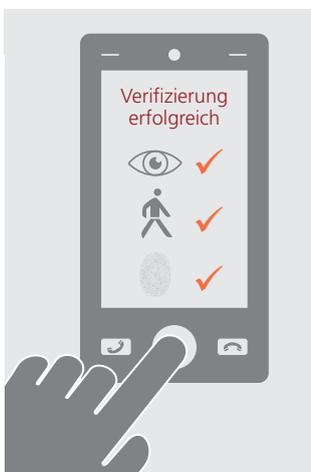

Für weiterführende Handlungen, die mit einem biometrisch abgesicherten Endgerät durchgeführt werden, bleibt allerdings zu berücksichtigen, dass dem Nutzer immer auch die Konsequenzen seines Handelns bewusst sein müssen. Führt er beispielsweise einen Kaufvorgang durch (z. B. ein mobiles Bezahlen) kommt der expliziten Bestätigung auch eine Signal- bzw. Warnfunktion zu.

Das Auflegen des Fingers, gegebenenfalls verbunden mit einer weiteren spezifischen Interaktionsmetapher, stellt einen naheliegenden Lösungsansatz dar. Eine weitergehende Betrachtung erfordert jedoch auch in diesem Gebiet übergreifende Forschungsbemühungen auf den Gebieten der Biometrie und der mobilen Mensch-Computer-Interaktion.

## Multifaktor-Authentifizierung

Bestimmte Anwendungsgebiete, wie z. B. das Online Banking, erfordern erhöhte Sicherheitsniveaus oder aber auch explizit eine Multifaktor-Authentifizierung. Biometrische Verfahren können mit wissensbasierten Verfahren kombiniert werden, beispielsweise indem bei der Eingabe im Falle des Fingerabdruckes auch gleichzeitig ein bestimmtes Entsperrmuster einzugeben ist. Dafür müssten allerdings zunächst die Forschungs- und Entwicklungsbemühungen im Bereich der Sensortechnologien verstärkt werden – denn das gesamte Display müsste in diesem Fall zur biometrischen Erfassung in der Lage sein.

Eine weitere vielversprechende Entwicklungslinie liegt in der Sprachbiometrie. Neben der Erkennung des Sprechers auf Basis

---

7 Busch, C. (2014): Biometrische Zugangskontrolle mit Smartphones. In: Datenschutz und Datensicherheit, Vol. 38 (7), S. 475–481.

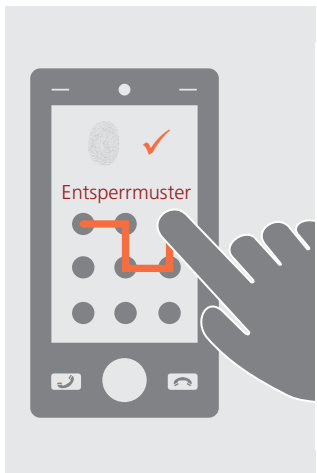

seiner sprachlichen Charakteristika kann auch hier ein Wissensfaktor, beispielsweise das Aussprechen eines Passwortes, miteinbezogen werden. Diese Funktionalität in Verbindung mit einer privatsphärenfreundlichen, ausschließlich lokalen Speicherung der biometrischen Referenzdaten auf mobilen Endgeräten alltagstauglich umzusetzen, stellt eine weitere große Forschungsherausforderung dar.

### Weiteres Anwendungspotenzial

Neben der technischen Frage, „wie" sich biometrische Verfahren sicher ein- und umsetzen lassen, stellt sich auch die Frage, „wozu" biometrische Ansätze vorteilhaft verwendet werden können. Die Nutzung biometrischer Verifizierungsverfahren als Ersatz für wissensbasierte Authentifizierungsverfahren wie Passwörter stellt zunächst nur den Ausgangspunkt für die Schaffung alltagstauglicher Sicherheitsverfahren dar. Die geeigneten „Killerapplikationen" zu identifizieren bleibt für die wirtschaftliche Verwertung von großer Bedeutung, sowohl kurz- als auch langfristig. Aus der Betrachtung des Marktes für Biometrielösungen lässt sich schließen, dass nicht eine einzelne Anwendung der Technologie zum endgültigen Durchbruch verhelfen wird.

Es zeichnen sich vielmehr verschiedene Bereiche ab, die in der Summe ein sehr großes Potenzial aufweisen. So könnten durch die Anbindung an existierende Vertrauensinfrastrukturen – wie etwa an elektronische Gesundheitskarten – bereits vielfältige Anwendungsfelder abgedeckt werden. So ist es zum Beispiel im Bereich des mobilen Bezahlens erforderlich, sichere und zugleich benutzerfreundliche Bezahlmöglichkeiten anzubieten. In welcher Form sich die bislang eher zurückhaltende Finanzindustrie auf die Nutzung biometrischer Lösung einlassen wird, bleibt jedoch noch abzuwarten.[8]

Zurzeit entstehen vielfältige „Pay-per-Use"-Modelle für Gebrauchsgüter, zu denen u. a. auch das weitverbreitete Carsharing zählt. Die Nutzung eines biometrischen Verfahrens kann hier den Zeitaufwand für das Beschaffen physikalischer Schlüssel deutlich verringern. Mobile Endgeräte werden zusehends mit biometrischen Erkennungsverfahren ausgestattet, die sich in absehbarer Zeit zu universell einsetzbaren Zugangsschlüssel weiterentwickeln können. Gegenüber reinen physikalischen Verfahren bietet ein solcher Ansatz einen weiteren entscheidenden Vorteil: Selbst wenn ein mobiles biometrisches Endgerät (und damit ein Schlüssel) verloren oder gestohlen wird, kann ein unehrlicher Finder es nicht nutzen. Somit können auch nachgelagerte Prozesse wie das Risikomanagement effizient gestaltet werden.

### Schlussfolgerungen

Es wird schon seit einiger Zeit kontrovers diskutiert, ob Biometrie die Souveränität des einzelnen Nutzers stärkt oder ob sie diese fundamental untergräbt. Zu den weit verbreiteten Vorbehalten gegenüber biometrischen Ansätzen gehören die pauschale Gleichsetzung mit Überwachungstechnologien sowie die damit verbundenen Assoziationen zur Strafverfolgung. Es ist jedoch wichtig, zwischen den Realisierungsvarianten biometrischer Systeme zu unterscheiden. Die Sorgen sind bei geeigneter Realisierung der Verfahren – und einem Einsatz zum alleinigen Zweck der Verifizierung – in vielen Fällen unbegründet.

Hier ist jedoch zuallererst ein Umdenken der Nutzerinnen und Nutzer notwendig, es sind psychologische Hürden zu überwinden – und das Vertrauen in die neuen biometrischen Technologien muss entscheidend gestärkt werden. Eine Durchführung von Datenschutzzertifizierungen nach anerkannten Standards kann dabei als vertrauensbildende Maßnahme dienen und den vertrauenswürdigen Umgang mit personenbezogenen Daten innerhalb biometrischer Systeme nachweisen.

Denn verbleiben personenbezogene biometrische Daten im Kontrollbereich der Nutzer, stellen biometrische Ansätze keine Überwachungstechnologien dar. Sie haben vielmehr das Potenzial, zu wirksamen Werkzeugen für den Schutz der Privatsphäre in der digitalen Welt zu werden, die sich auch von technischen Laien komfortabel nutzen lassen. Damit weisen biometrische Verfahren große Vorteile im Bereich der Benutzbarkeit von IT-Sicherheits- und Datenschutzmechanismen auf und könnten somit als flexibler Ersatz für bisherige Authentifizierungsverfahren eingesetzt werden.

Die Verbindung von Biometrie mit mobilen Endgeräten ist dabei einer der wichtigsten Entwicklungstrends. Perspektivisch lassen sich auf diese Weise u. a. zuverlässige und einfach zu benutzende Universalschlüssel für das Internet der Dinge und Dienste realisieren.

---

Es bleibt jedoch festzustellen, dass für den erfolgreichen Einsatz von Biometrie noch weitere Forschungs- und Entwicklungsbemühungen notwendig sind. So zum Beispiel

- die Entwicklung von Sensoren, die eine parametrisierte, großflächige Erfassung sowie eine zuverlässigen Manipulationserkennung bieten,
- die Fusion biometrischer Merkmale und Modalitäten sowie
- die Realisierung geeigneter Hardware- und Softwareplattformen für mobile Endgeräte, um biometrische Referenzdaten geschützt im Kontrollbereich der Nutzer vorhalten zu können.

Es zeigt sich, dass auch die Berücksichtigung von Designaspekten einen zentralen Akzeptanzfaktor für biometrische Authentifizierungslösungen darstellt. Aus einer übergreifenden Betrachtung der Gebiete der Biometrie und der mobilen Mensch-Computer-Interaktion hat sich bereits ein sehr aktiver Forschungsbereich entwickelt, der durch das Eintreten der Biometrie in den Massenmarkt einen zusätzlichen Schub erhalten hat.

Werfen wir abschließend noch einmal einen Blick zurück: Der technologische Fortschritt der letzten zwei Dekaden hat effiziente biometrische Verifizierungsverfahren hervorgebracht und deren datenschutzgerechte Umsetzung ermöglicht. Der deutsche Gesetzgeber hat das große Anwendungspotenzial der besprochenen Technologien frühzeitig erkannt und biometrische Verfahren schon zu Anfang des Jahrtausends als Realisierungsvariante für digitale Identitätsnachweise zugelassen. Biometrische Verfahren, die mit mobilen Endgeräten gekoppelt sind, stellen nun eine wirksame Alternativen zu wissensbasierten Authentifizierungsverfahren wie Passwort und PIN dar – die nun in vielfältigen Erscheinungsformen auf den Massenmarkt drängen.

Die erfolgreiche Vermarktung hat somit gerade erst begonnen. Die deutsche und europäische Wirtschaft hat hier vielfältige Potenziale und gute Chancen, künftig weltweit eine führende Rolle einzunehmen.